\documentclass[twocolumn]{aastex62}

\usepackage{float}

\newcommand{\oiii}{[O~{\footnotesize III}]}
\newcommand{\nii}{[N~{\footnotesize II}]}
\newcommand{\sii}{[S~{\footnotesize II}]}
\newcommand{\siii}{[S~{\footnotesize III}]}
\newcommand{\neiii}{[Ne~{\footnotesize III}]}
\newcommand{\cloudy}{C{\footnotesize LOUDY}}

\def\etal{{et~al.\null}}

\def\Gaia{{\it Gaia}}
\def\HST{{\it HST}}

\def\bigstar{{\raisebox{-1.8pt}{\large$\ast$}}}

\graphicspath{{./}{figures/}}

\submitjournal{ApJ}

\shorttitle{Planetary Nebula in an Open Cluster in M31}

\shortauthors{Davis \etal}

\begin{document}


\title{{\it Hubble Space Telescope\/} Spectroscopy of a Planetary Nebula in an M31 Open Cluster: \\ Hot-Bottom Burning at $3.4\, M_\odot$\footnote{Based on observations with the NASA/ESA {\it Hubble
Space Telescope\/} obtained at Space Telescope Science Institute, operated by
Association of Universities for Research in Astronomy, Inc., under NASA contract
NAS5-26555.}}



\correspondingauthor{Brian D. Davis}
\email{bdavis@psu.edu}

\author[0000-0002-8994-6489]{Brian D. Davis}
\affil{Department of Astronomy \& Astrophysics, The Pennsylvania
State University, University Park, PA 16802, USA}

\author[0000-0003-1377-7145]{Howard E. Bond}
\affil{Department of Astronomy \& Astrophysics, The Pennsylvania
State University, University Park, PA 16802, USA}
\affil{Space Telescope Science Institute, 3700 San Martin Dr., Baltimore, MD 21218, USA}

\author[0000-0002-1328-0211]{Robin Ciardullo}
\affil{Department of Astronomy \& Astrophysics,
The Pennsylvania
State University, University Park, PA 16802, USA}
\affil{Institute for Gravitation and the Cosmos, The Pennsylvania
State University, University Park, PA 16802, USA}

\author[0000-0001-7970-0277]{George H. Jacoby}
\affil{Lowell Observatory, Flagstaff, AZ 86001, USA}



\begin{abstract}

We use imaging and spectroscopy from the {\sl Hubble Space Telescope\/} (\HST) to examine the properties of a bright planetary nebula (PN) projected within M31's young open cluster B477-D075.  We show that the probability of a chance  superposition of the PN on the cluster is small, ${\lesssim}2\%$. Moreover, the radial velocity of the PN is the same as that of the cluster within the measurement error of $\sim\!10$~km~s$^{-1}$.  Given the expected $\sim\!70$~km~s$^{-1}$ velocity dispersion in this region, ${\sim}$8~kpc from M31's nucleus, the velocity data again make it extremely likely that the PN belongs to the cluster.  Applying isochrone fitting to archival color-magnitude photometric data from the \HST\/ Advanced Camera for Surveys, we determine the cluster age and metallicity to be 290~Myr and $Z = 0.0071$, respectively, implying an initial mass of $3.38^{+0.03}_{-0.02} \, M_{\odot}$ for any PN produced by the cluster. From \HST's Space Telescope Imaging Spectrograph observations and \cloudy\ photoionization modeling, we find that the PN is likely a Type~I planetary, with a nitrogen abundance that is enhanced by ${\sim}$5--6 times over the solar value scaled to the cluster metallicity. If the PN is indeed a cluster member, these data present strong empirical evidence that hot-bottom burning occurs in AGB stars with initial masses as low as $3.4 \, M_{\odot}$.


\end{abstract}


\keywords{planetary nebulae: general --- galaxies: individual (M31) --- galaxies: stellar content --- stars: AGB and post-AGB}


\section{Planetary Nebulae in Star Clusters}
\label{sec:intro}

Planetary nebulae (PNe) mark the spectacular transitions of low- and
intermediate-mass (${\sim}$1--$8 \, M_{\odot}$) stars from the asymptotic giant
branch (AGB) to the onset of the white-dwarf cooling sequence.  Objects in this short-lived phase are critically important to our understanding of stellar evolution, as they allow us to test the physics of nucleosynthesis, the systematics of dredge-up, and the relation between the initial mass of a star and the mass of its white-dwarf
remnant. Unfortunately, astrophysical studies of PNe are hampered by the fact
that, even with the recent release of {\it Gaia\/} parallaxes, the distances to
most Galactic PNe remain uncertain. Moreover, for nearly all PNe, the masses and compositions of the progenitor stars are unknown, or at best can only be roughly
inferred from statistical studies of the underlying stellar population \citep[e.g.,][]{badenes2015}.

The rare PNe that are members of star clusters provide important exceptions to
the uncertainties associated with distance and origin. For these objects,
the progenitor star's mass and metallicity can be determined from the cluster's
main-sequence turnoff and metal content, the PN's absolute luminosity
can be found using the cluster's distance via main-sequence
fitting (or membership in an external galaxy), and the mass of the PN central star can be deduced from theoretical core-mass\slash
luminosity relations. Moreover, the chemical composition of the PN can provide 
information about the dredge-up of chemical elements, which is expected to depend on the initial mass of the star.

There are a more than a dozen known cases of PNe lying near Galactic open
clusters. The extensive literature on these putative PN-cluster associations has been summarized by
\citet{Majaess2007}, \citet{Bonatto2008}, \citet{MoniBidin2014},
\citet{Frew2016}, \citet{Gonzalez2019}, and references therein. Nearly all of these objects are chance
superpositions; \citet{Frew2016} found only a handful of the associations
convincing enough to use as primary PN distance calibrators.

The difficulty of definitively associating Milky Way PNe with Galactic open clusters is illustrated by two well-known
cases. (1)~The most famous is the bright PN NGC\,2438,
which is conspicuous in the field of the open cluster M46. Studies of these two
objects date back over 70 years, with \citet{Cuffey1941} and \citet{Odell1963}
being among the first to question their physical association. Although
\citet{Pauls1996} argued that the PN is a cluster member, a more
recent radial-velocity (RV) study \citep{Kiss2008} casts doubt on this
conclusion.   In any case, an association is now conclusively ruled
out, as the \Gaia\/ Data Release~2 (DR2) central-star parallax and proper motion \citep{Gaia2018, Kimeswenger2018} differ significantly from
those of cluster members. (2)~The PN NGC\,2818 lies close to an open cluster
with the same designation \citep{Tifft1972}, and RV measurements
\citep{Vazquez2012} suggest a physical association. Moreover, based on its RV,
H$\alpha$ surface brightness, and radius, \citet{Frew2016} concluded that the PN
is indeed a cluster member. However, the \Gaia\/ DR2 data again rule out this
possibility, as the proper motion of the PN central star is clearly discordant
with those of cluster members \citep{Gaia2018, Bastian2018}.

At present, such considerations leave only two likely cases of a PN
belonging to a Milky Way open cluster: (1)~PHR~J1315$-$6555 is probably associated with
the open cluster ESO\,96-SC04 = Andrews-Lindsay\,1
\citep{Parker2011, Majaess2014}. In a follow-up analysis,
\citet{Fragkou2019b} determined the age (660~Myr) and turnoff mass of the cluster
(${\sim}2.2\,M_\odot$), the mass of the PN central star ($0.58\,M_\odot$), and
other astrophysical parameters associated with the PN and cluster.
(2)~BMP\,J1613$-$5406 was reported to be a likely member of the open cluster NGC\,6067 by
\citet{Frew2016}. A recent study of the PN and cluster by \citet{Fragkou2019a} found an age of 90~Myr for the host cluster and a PN progenitor mass of $\sim\!5.6\,M_\odot$. The mass of the PN central star was inferred to be about $0.94\,M_\odot$, and the nebula appears to have an enhanced nitrogen abundance.

Turning to the Milky Way's ancient globular clusters (GCs), we note that four are known to contain PNe 
\citep{Pease1928, Gillett1989, Jacoby1997}.  However, the stars currently evolving in GCs have
masses of about $0.8\,M_\odot$, implying extremely long evolutionary timescales for their post-AGB remnants; the ejected nebulae from these objects will almost certainly disperse before ionization can occur. The PNe that {\it are\/} seen within GCs are most likely descended from interacting or merged
binaries, telling us little about normal single-star evolution \citep[e.g.,][hereafter B15]{Alves2000, Jacoby2013, Jacoby2017, Bond2015}. 

Two recent surveys have been carried out for PNe projected upon the star clusters of nearby Local Group galaxies.\footnote{It is even more difficult to search for PNe associated with star clusters at distances beyond the Local Group. However, a PN within a GC in NGC\,5128 (Centaurus~A, distance $\sim$3.7~Mpc) was reported by \citet{Minniti2002}, and \citet{Larsen2006} found three PN candidates superposed upon young open clusters in M83 (distance 4.5~Mpc) and  NGC 3621 (6.6~Mpc).}
One searched for [\ion{O}{3}] 5007~\AA\ emission within M31 clusters using ground-based spectra of the clusters' integrated starlight \citep{Jacoby2013}; the other used 
{\it Hubble Space Telescope\/} (\HST) direct imaging through a narrow-band [\ion{O}{3}] filter to examine star clusters throughout the Local Group
(B15). Most of the targeted clusters in these two programs were GCs, but a few open clusters were also included. These studies have produced a number of candidate associations, most of which require
follow-up investigations for confirmation. However, one  association appears unambiguous: a bright PN located within a young M31 open cluster.
This PN and its host cluster are the subjects of this paper.

\section{A Planetary Nebula in an Open Star Cluster in M31}

In the course of an investigation of archival narrow-band [\ion{O}{3}] \HST\/
images of fields in M31, B15 noted a previously unrecognized case of a
conspicuous PN candidate projected within a known star cluster. The cluster
is listed as B477 (or Bol\,477) in two catalogs of M31 GCs: the Bologna
Catalogue\footnote{{\url{http://www.bo.astro.it/M31}}; Catalogue Version~5, 2012
August} \citep{Galleti2004}, and \citet{Peacock2010}. In the naming scheme
recommended by \citet{Barmby2000}, the cluster is designated B477-D075, with the ``D'' referring to the earlier ``DAO'' listing of candidate M31 GCs by \citet{Crampton1985}. We use this name for the cluster hereafter. The PN was given the designation M31~B477-1 by B15, and its J2000 position, 00:45:08.42,
+41:39:38.4, lies $1\farcs2$ northeast of the cluster center, which is at 00:45:08.33, +41:39:38.0.  
This location falls in the disk of M31 on the northeast side of the nucleus, near the galaxy's major axis. Its angular distance from the galaxy center is $35\farcm8$, which, for an M31 distance of 750~kpc \citep{Riess2012}, corresponds to a projected distance of 7.8~kpc.  


\subsection{The Cluster}
\label{sec:cluster}
Archival \HST\/ images that serendipitously cover the location of B477-D075 are available\footnote{\url{http://archive.stsci.edu/}} from two programs. In 2003, \HST\/ proposal
GO-9794 (PI: P.~Massey) used the Advanced Camera for Surveys (ACS) to image the region in the
broad-band $B$ (F435W), $V$ (F555W), $R$~(F625W), and $I$ (F814W) bandpasses,
and the narrow-band [\ion{O}{3}] (F502N) and H$\alpha$+[\ion{N}{2}] (F658N) filters.
During 2011--2013, the site was imaged again as part of the 
Panchromatic Hubble Andromeda Treasury (PHAT)
project \citep[GO-12071, \hbox{-12111}, -12114, and -12115;][]{Dalcanton2012}.  The PHAT data consist of a variety
of broad-band images obtained with near-ultraviolet, optical, and near-infrared filters, using both the ACS and the Wide
Field Camera~3 (WFC3). 

\begin{figure}
\centering
\includegraphics[width=0.473\textwidth]{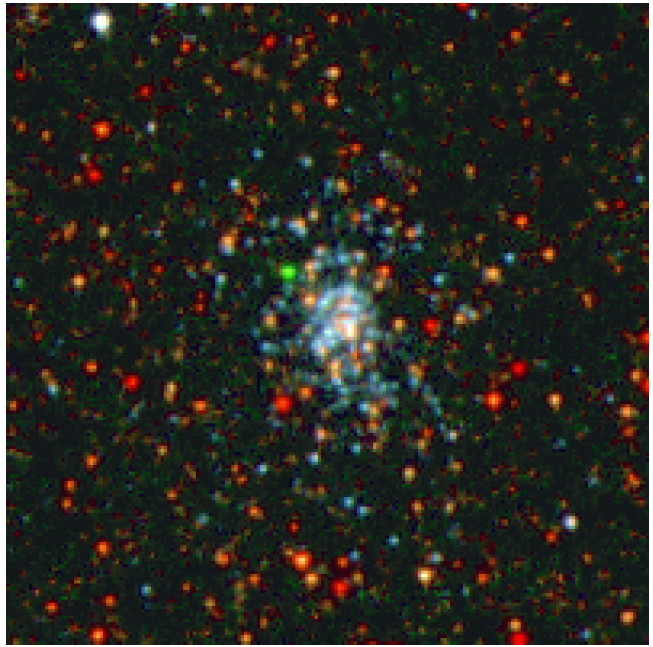}
\caption{Color rendition of the M31 cluster B477-D075 and its PN,
created from archival \HST/ACS images in F435W (blue), F502N+F555W (green), and
F814W (red). The frame is $10''$ high, with north roughly at the top and east on the
left. The PN, whose [\ion{O}{3}] 5007~\AA\ emission is detected in the F502N and F555W images, 
is seen as a bright green point source northeast of the cluster
center. The morphology and predominance of blue stars show B477-D075 to be an open cluster. Cluster members extend well beyond the position of the PN.
}
\label{fig:prettypic}
\end{figure}

Figure~\ref{fig:prettypic} shows a color image of B477-D075, created from the
$B$, $V$+[\ion{O}{3}], and $I$ frames of the GO-9794 ACS program; the blue,
green, and red colors were assigned as described in the caption.
The PN stands out as the bright green point source lying northeast of the
cluster's center. 

Although B477-D075 had been listed in several catalogs of candidate M31 GCs, the morphology and predominance of blue stars seen in
Figure~\ref{fig:prettypic} clearly show it to be an open cluster. As such, it has been
included in several investigations of young star clusters in M31.   \citet{FusiPecci2005} listed the system as a possible young cluster, based on its blue color, but provided no other information. \citet[][hereafter C09; their Fig.~4]{Caldwell2009} published a grayscale image of the cluster from the
same ACS $V$ frame that we use here, and likewise classified it as a young open
cluster. Based on optical integrated-light spectra, they obtained an age of about
316~Myr, an RV of $-110.4\pm12\,\rm km\,s^{-1}$, and a
reddening of $E(B-V)=0.30$. C09 also noted ``emission'' in the integrated spectrum,
without further comment. (In a recent private communication, N.~Caldwell informed us that their quoted RV was based on the wavelengths of the emission lines.)

\citet{Kang2012} obtained ultraviolet (UV) measurements of the cluster and used the system's full far-UV to near-infrared spectral-energy distribution (SED) to derive a
smaller cluster reddening of $E(B-V)=0.08$, but nearly the same age of 325~Myr. However,
a similar SED analysis by \citet{Wang2012} gave an age of only 62~Myr, and a
reddening of $E(B-V)=0.30$. \citet{Chen2015} independently determined the
cluster's RV using the Large Sky Area Multi-Object Fiber Spectroscopic Telescope
(LAMOST), finding $-123\pm12~\rm km\,s^{-1}$. The LAMOST group then performed a population-synthesis analysis of the
integrated-light spectrum and derived a cluster metallicity of $Z=0.0067$ 
\citep{Chen2016}.

According to the kinematic model of \citet{Braun1991}, objects belonging to M31's disk at the position of the cluster should have a heliocentric RV of $-113\,\rm km\,s^{-1}$ for the most likely spiral-arm membership, or $-160\,\rm km\,s^{-1}$ for a second possible arm membership \citep{Walterbos1992}. M31 disk stars at the location of the cluster have a velocity dispersion of between ${\sim}$40 and ${\sim}$70~km~s$^{-1}$, depending upon their age \citep{Merrett2006, Dorman2015}.
Thus the measured RVs of the cluster are consistent with membership in the disk population.

By comparing broad-band integrated \HST\/ photometry published by the PHAT project to a grid of theoretical models, \citet{deMeulenaer2017}
inferred for the cluster an extinction of $E(B-V)=0.07$, a metallicity of $\rm[Fe/H]=+0.2$, and an
age of 250~Myr. The PHAT team themselves \citep{Johnson2016} determined an
age of 200~Myr, based on photometry of the cluster's stars as seen in their \HST\/
images, and isochrone fitting of the resultant color-magnitude diagram (CMD\null).

Table~\ref{table:clusterdata} gives a summary of the cluster parameters obtained from the existing studies of B477-D075. The previous work has yielded cluster ages
ranging from about 200 to 325~Myr, except for one analysis that gave a much lower value,
close to 60~Myr. Fairly wide ranges of reddenings and cluster metallicities
have also been found. We will make our own age, reddening, and metallicity
determinations later in this paper, based on an analysis of the cluster's CMD and using model stellar isochrones. Anticipating our results given below, we list these values in the final row of Table~\ref{table:clusterdata}.

\begin{deluxetable*}{lcccl}
\tablewidth{0pt}
\tablecaption{Properties of M31 Cluster B477-D075\label{table:clusterdata}}
\tablehead{
\colhead{Age}        &
\colhead{$E(B-V)$}        &
\colhead{[Fe/H]}        &
\colhead{Radial Vel.}        &        
\colhead{Reference}\\
\colhead{[Myr]}        &
\colhead{[mag]}        &
\colhead{[dex]}             &
\colhead{[$\rm km\,s^{-1}$]} & 
\colhead{}        
}
\startdata
%
%
316	& 0.30         &  $\dots$	 & $-110.4\pm12$ &  \citet{Caldwell2009} \\ 
325     & 0.08         &  $\dots$	 & $\dots$	 &  \citet{Kang2012}	 \\
$62\pm6$& $0.30\pm0.05$&   $\dots$	 & $\dots$	 &  \citet{Wang2012}	 \\
$\dots$ & $\dots$      &  $\dots$	 & $-123\pm12$   &  \citet{Chen2015}	 \\
$\dots$ & $\dots$      &  $-0.33$\tablenotemark{a}	 & $\dots$	 &  \citet{Chen2016}	 \\
200     & $\dots$      &  $\dots$	 & $\dots$	 &  \citet{Johnson2016}  \\
250     & 0.07         & +0.2 & $\dots$	 &  \citet{deMeulenaer2017}\\
\noalign{\smallskip}
$290\pm5$ & $0.171\pm0.003$ & $-0.30\pm0.02$\tablenotemark{a} & $\dots$   & This work \\
\enddata
\tablenotetext{a}{Reference gives mass fraction of metals, $Z$, which we convert to [Fe/H] using $Z_\odot=0.0142$.}
\end{deluxetable*}

\subsection{The Planetary Nebula}

The PN superposed near B477-D075 was first cataloged as an H$\alpha$ emission-line source by
\citet[][their object no.~683]{Walterbos1992}, who noted that it appeared to
coincide with a continuum source (in retrospect, the star cluster). 
Subsequently, during a survey of M31 with the Planetary Nebula Spectrograph (PN.S), the object
was identified as a candidate PN by \citet[][hereafter M06; their object
no.~446]{Merrett2006}. These authors measured an [\ion{O}{3}] $\lambda 5007$ magnitude for 
the PN of $m_{5007}=-2.5\log(F_{5007})-13.74=22.30$, where $F_{5007}$ is the monochromatic flux in the [\ion{O}{3}] emission line in erg~cm$^{-2}$~s$^{-1}$. They also estimated the PN's RV to be $-150.1\pm14\,\rm km~s^{-1}$.

In a study of broad-band \HST\/ images by the PHAT team \citep{Veyette2014},
the emission-line object was noted as appearing extended, and was thus classified as an \ion{H}{2} region, rather
than an unresolved PN\null. In retrospect, we see that the extended object is
the host star cluster, not the emission-line object itself.

\section{Is the PN a Cluster Member?}\label{sec:isitamember}

In general, there are five criteria for evaluating whether a PN can be considered a likely member of a star cluster: (1)~the PN must be superposed on the sky within the boundaries of the cluster; (2)~the RV of the PN must be consistent with that of the cluster and the RV dispersion of its members; (3)~the reddening of the PN and its central star should be at least as large as that of the cluster (but could be larger due to circumnebular extinction); (4)~the distance of the PN should be the same as that of the cluster, to within the uncertainties; and (5)~the proper motion of the PN's central star should be consistent with those of the cluster stars.

In principle, all five criteria can be applied in the Milky Way. However, the equal-distance requirement is often indecisive due to the large uncertainties in PN distance estimates. Additionally, in the Milky Way, the projected angular separation between a PN and a star cluster carries little information: not only is the likelihood of a chance superposition high \citep{Frew2016}, but our location within the Galaxy makes estimating the precise probability of such an occurrence extremely difficult to calculate. More specifically, the large numbers of known Galactic PNe and clusters, and the large angular radii of the clusters create many spurious positional superpositions. As we noted in \S1, there are only two cases of Galactic PNe to date that have passed the stringent requirements for membership in open clusters.

When we survey extragalactic PNe, our position outside the system makes the situation quite different. Not only is it relatively easy to compute the non-posterior probability of a PN-cluster superposition, but, in general, the likelihood of such coincidences is quite low, due to the limited number of PNe bright enough to be detected, and the small angular sizes of the star clusters. Consequently, although one cannot test for agreement in distance or proper motion, projected angular separation becomes a powerful tool for assessing the probability of association. In this section, we perform statistical tests of the superposition and RV membership criteria for our target system. The reddening is discussed in \S6.2.

\subsection{Probability of Superposition}

The effective angular radius of B477-D075 (i.e., the cluster's half-light radius) is given by \citet{Johnson2015} as $0\farcs99$. (This is only an approximation, since the cluster is noticeably elongated.)  As can be seen in Figure~\ref{fig:prettypic}, the PN lies just outside the effective radius, at a distance of $1\farcs2$ from the cluster center, and nearly along the system's long axis. There are obvious blue cluster members at larger separations. 

We estimated the probability of a chance alignment from two different standpoints. First we defined two large
regions of interest: one $950 \, \rm{arcmin}^2$ field lying in the northeast
disk of M31 containing B477-D075 and a representative disk population ($ 11\fdg535
> \alpha > 11\fdg035$,  $ +41\fdg419 < \delta < +42\fdg119$), and a region of the
same size on the symmetrically opposite southwest side of the nucleus ($ 10\fdg335 >
\alpha > 9\fdg835$, $ +40\fdg319 < \delta < +41\fdg019$). Both areas have
existing deep surveys for both PNe and star clusters.

These two fields contain a total of 169 star clusters in the C09 catalog, of which 102 are of ``young'' ($<\!1\,\rm Gyr$) or
``intermediate'' (1--2~Gyr) age \citep[including 24 clusters reclassified from
``old'' by][]{Caldwell2011}. We take the probability of a single PN lying within $1\farcs2$ of a cluster to be the ratio of the total areas subtended by the clusters to the total $1900 \, \rm{arcmin}^2$ area of the two regions; these are 0.0112\% for all clusters, and 0.00674\% for the young- and intermediate-age clusters. In the same two regions there are 167 PNe with
$m_{5007} \le22.3$ in the M06 catalog. These data yield the
probability of finding a PN this bright by chance within $1\farcs2$ of any cluster as 1.9\%.  This probability drops to 1.1\% if we only consider the young- to intermediate-age clusters.

Alternatively, we consider the specific ACS [\ion{O}{3}] frame from \HST\/ program
GO-9794 on which B15 noted the apparent association of a PN with the
cluster B477-D075. This frame covers an area of about $11.6\,\rm arcmin^2$, and
it contains four star clusters from the C09 catalog. Thus the fractional area of
the frame lying within $1\farcs2$ of a cluster is 0.043\%. We searched 
the M06 catalog for PNe that are bright enough ($m_{5007}\la24.5$) to
detect in this frame, finding seven. This leads to a probability of finding one
of them by chance within $1\farcs2$ of a cluster of only 0.30\%. Thus, the location of the PN within the cluster strongly suggests membership.

\subsection{Probability of Radial-Velocity Agreement} 
The cluster B477-D075 has two RV measurements in the literature:  $-110 \pm 12$~km\,s$^{-1}$ from C09 and $-123 \pm 12$ km\,s$^{-1}$ from the LAMOST observations of \citet{Chen2015}.  The PN has only one velocity determination:  $-150.1 \pm 14$~km\,s$^{-1}$ from the PN.S M31 survey (M06). On their face, these two values are in only marginal agreement.  However, the C09 velocity for the cluster included information from the cluster's apparent emission lines, while the PN.S RVs are known to be susceptible to systematic offsets (M06).  Thus, the question deserves further investigation.

The spectrum discussed by C09 was obtained using a $1\farcs 5$-diameter fiber of the MMT's Hectospec spectrograph under conditions with roughly $1\farcs 5$ seeing. As C09 noted, both the absorption-line
integrated-light spectrum of the cluster, and the emission lines from the nebula, are
present in their spectrum.\footnote{Although C09 did not publish
the spectrum, a plot is available at \url{https://www.cfa.harvard.edu/oir/eg/m31clusters/B477-D075\_y.html}} N.~Caldwell
kindly provided us with this spectrum, allowing us to measure the relative wavelengths of these lines.  We find the difference between the velocity determined from the PN's five brightest emission lines and that determined from the cluster's six cleanest absorption features to be 
$+4.5\pm10.7\,\rm km\,s^{-1}$, i.e., consistent with both objects having the
same RV\null.  For comparison, at an M31 disk radius of ${\sim}8$~kpc, M06 measured the line-of-sight velocity dispersion of PNe to be ${\sim}70$~km\,s$^{-1}$, while \citet{Dorman2015} obtained velocity dispersions of ${\sim}40$~km\,s$^{-1}$ and ${\sim}70$~km\,s$^{-1}$ for young- ($<$1~Gyr) and intermediate- ($<$4~Gyr) age AGB stars, respectively.  This implies that the likelihood of an unrelated PN having an RV within $\pm 5$~km\,s$^{-1}$ of B477-D075 is $\lesssim$5--10\%.  Consequently, although our RV measurements do not conclusively prove membership, in combination with the spatial coincidence, they make it extremely likely (${>}99.9$\%) that the PN is physically associated with the cluster. For the rest of this paper, we assume that the PN is a member of the star cluster.


\section{{\it HST\/} STIS Observations and Reductions} \label{sec:reduction}

\subsection{Observations}

With the aim of determining nebular parameters and chemical abundances, we obtained optical and UV spectroscopy of the PN M31~B477-1 with the
Space Telescope Imaging Spectrograph (STIS) onboard \HST, under program
GO-14794 (PI: H.E.B.). The high spatial resolution of \HST\/ allowed us to acquire spectra of
the PN free of contamination from the host star cluster; this is especially important for measurements of the Balmer emission lines, which would lie near the cores of the strong absorption features of the cluster's underlying A-type integrated-light spectrum.
Details of the observations are given in
Table~\ref{table:hstobs}. The detectors for the UV (gratings G140L and G230L) spectra were the STIS
multi-anode microchannel arrays (MAMAs); for the optical spectra (gratings G430L and G750L), data were recorded by a CCD\null.


\begin{deluxetable*}{lcccccc}
\tablewidth{0pt}
\tablecaption{\HST/STIS Observations of M31~B477-1\label{table:hstobs}}
\tablehead{
\colhead{Date}        &
\colhead{Detector}        &
\colhead{Grating}        &
\colhead{Wavelength}        &
\colhead{Dispersion}        &
\colhead{Resolving}        &
\colhead{Exposure}\\
\colhead{}        &
\colhead{}        &
\colhead{}        &
\colhead{Range [\AA]}        & 
\colhead{[\AA~$\rm pix^{-1}$]}        &
\colhead{Power [$\lambda/2\Delta\lambda$]} &
\colhead{[s]}
}
\startdata
2017 Nov.\ 29 & FUV-MAMA & G140L & 1150--1730  & 0.6 & $\sim\! 1200$ & $2\times1435$ \\
\qquad$''$    & NUV-MAMA & G230L & 1570--3180  & 1.6 & $\sim\! 750$ & $2\times1003$ \\
2017 July 25 & CCD      & G430L & 2900--5700  & 2.7 & $\sim\! 780$ & $4\times690$ \\
\qquad$''$    & CCD      & G750L & 5240--10270 & 4.9 & $\sim\! 780$ & $3\times615$ \\
\enddata
\end{deluxetable*}

All of our STIS observations were obtained using a long slit with angular dimensions of
$52''\times0\farcs5$. We selected this configuration to obtain the highest
throughput for our faint target, but since this slit width is larger than the full
width at half-maximum (FWHM) of the PN, our data do not allow for a precise RV measurement.
Because the PN is too faint for direct acquisition onto the STIS
slit, our observing strategy was to acquire 
a nearby bright star, and perform a blind offset onto the PN, using precise relative astrometry that we measured on the PHAT images.
For all of the observations, we performed small dithers along the slit between
each integration. For the optical observations, the target was placed near the edge
of the CCD detector, in order to minimize the effects of charge-transfer inefficiency.

\subsection{Data Reduction}

We reduced our spectroscopic observations using \texttt{calstis} \citep{Hodge1998}, the standard reduction package for STIS\null. We followed the reduction steps required by \texttt{calstis} to obtain calibrated 2D spectra of our object from the dithered data produced by the MAMA and CCD detectors. The results of these reductions are wavelength- and flux-calibrated 2D images (with pixel values in $\rm erg\,\, s^{-1}\, cm^{-2}\, {\mbox{\AA}}\null^{-1}\, arcsec^{-2}$).

Once these images were obtained, we modeled the point-spread function (PSF) of the 2D spectra, collapsed the spectra to 1D by co-adding the pixels within the PSF’s FWHM, and then scaled the resultant line fluxes to account
for flux outside the extraction aperture. In this way, the correct amount of flux could be measured
without introducing excess noise into the final spectrum.

\begin{figure*}
\centering
\includegraphics[width=\textwidth]{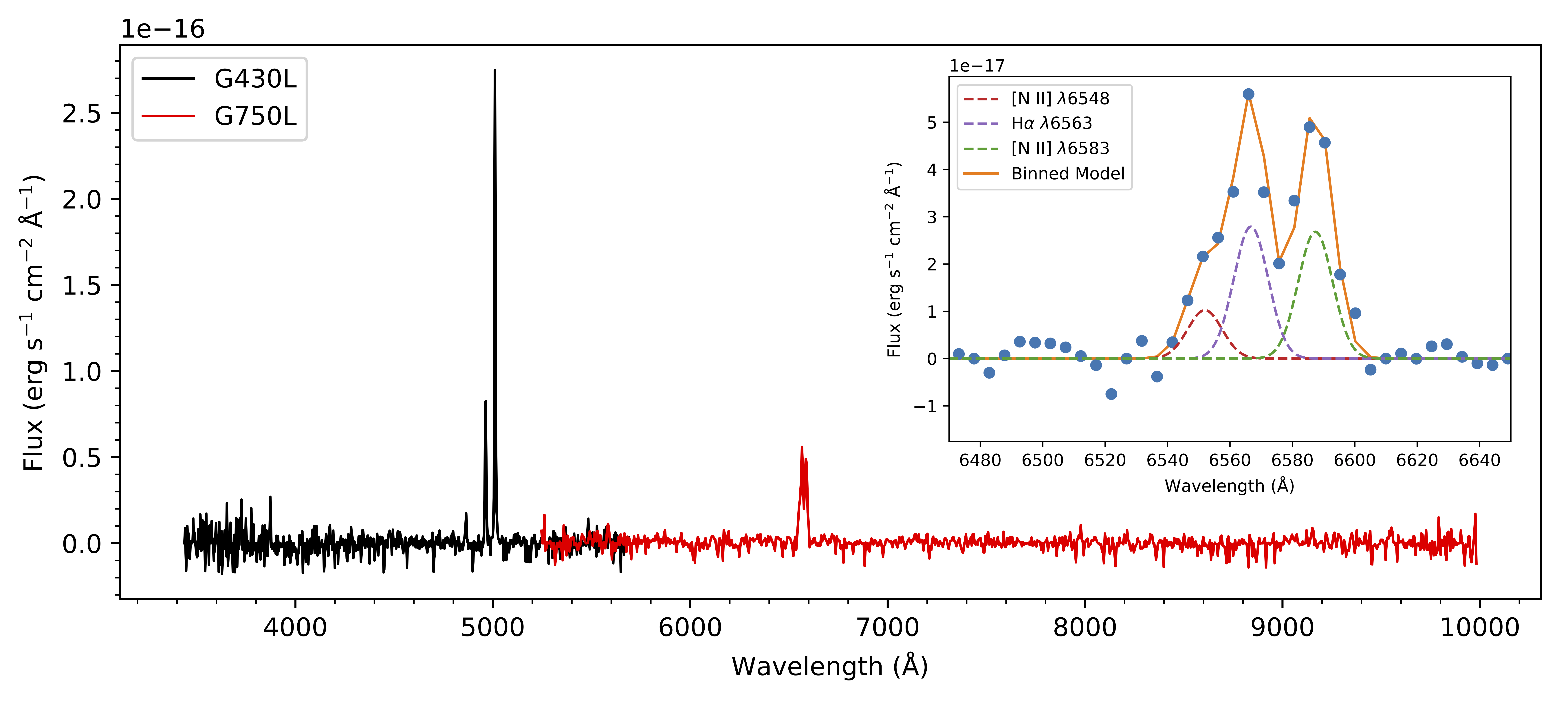}
\caption{Spectra of the PN in the open cluster B477-D075, obtained with the STIS G430L and G750L gratings, shown as black and red lines, respectively. The spectrum is dominated by emission from \oiii\ $\lambda\lambda 4959, 5007$, the [\ion{N}{2}] doublet, and, to a lesser extent, the Balmer lines. {\it Inset:} Our fit for the \nii\ $\lambda 6548$, H$\alpha$, and \nii\ $\lambda 6583$ complex in the G750L grating. Dashed lines show the modeled line profile of each emission line, scaled down by a factor of two for visibility. The solid orange curve shows the sum of the three individual emission-lines, binned to the spectrograph resolution. Note that the [\ion{O}{2}] $\lambda 3727$ emission line is, at best, barely distinguishable from the noise, but \nii\ $\lambda 6583$ is as bright as H$\alpha$. This indicates a large overabundance of nitrogen, consistent with a Type~I PN.}
\label{fig:full_spec}
\end{figure*}

Figure~\ref{fig:full_spec} shows the PN's STIS spectrum measured with the G430L (black line) and G750L (red line) gratings. This spectrum is typical of a PN that is within $\sim$2.5~mag of the bright-end cutoff of M31's PN luminosity function \citep{Ciardullo2002, Merrett2006, Bhattacharya2019}.  Specifically, \oiii\ $\lambda 5007$ has more than twice the 
strength of H$\alpha$+\nii\ \citep{Ciardullo2002}, the 
H$\alpha$/H$\beta$ ratio is much greater than the \citet{Baker1938} 
Case~B value of 2.85 \citep{Brocklehurst1971, Davis2018}, and the excitation level, as measured by the strength of \oiii\ $\lambda 5007$ relative to H$\beta$, is high \citep[e.g.,][]{Mendez2005, Herrmann2009, Reid2010}. Unfortunately, the MAMA spectra showed no
significant line detections in the UV, with upper limits for \ion{C}{4} $\lambda 1550$ and \ion{C}{3}] $\lambda 1909$ of 
${\sim}10^{-16}$~erg~cm$^{-2}$~s$^{-1}$ (about 50-70\% of H$\beta$).

\begin{figure}
\centering
\includegraphics[width=0.475\textwidth]{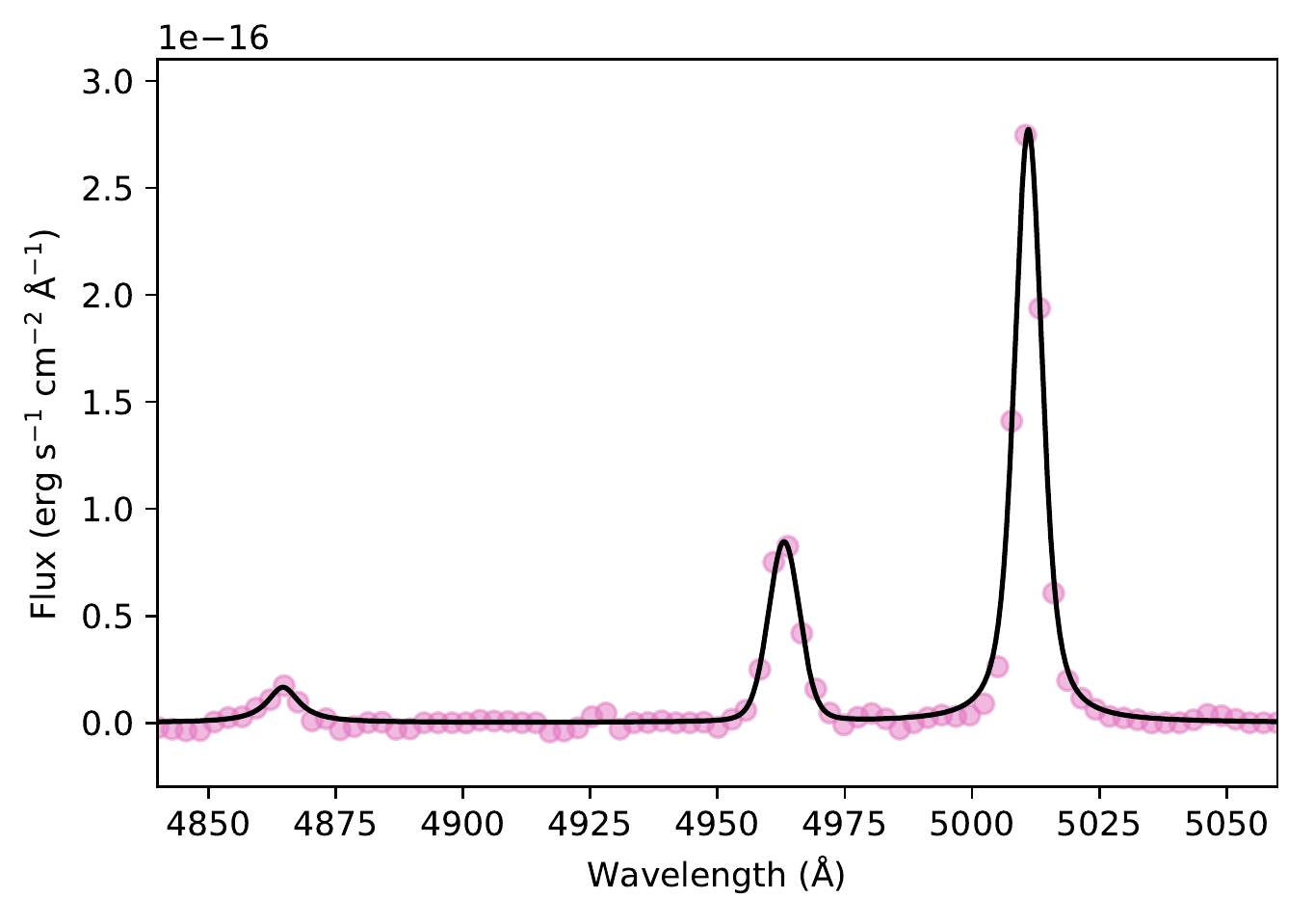}
\caption{The pink points show a segment of the PN's STIS G430L spectrum near the emission lines of H$\beta$, \oiii\ $\lambda 4959$, and \oiii\ $\lambda 5007$.  The black line illustrates our Gaussian fits to these lines.  The \oiii\ $\lambda 5007$/H$\beta$ ratio of ${\sim}$15 is typical of a PN in the top ${\sim}$2~mag of the PN luminosity function.}
\label{fig:G430L}
\end{figure}

To obtain monochromatic line fluxes from the spectra, we modeled each line in Table~\ref{table:lines} using a Gaussian profile on top of a background continuum.  For isolated lines such as H$\beta$ and \oiii\ $\lambda\lambda 4959, 5007$, this procedure was straightforward (see Figure~\ref{fig:G430L}), as the continuum is extremely weak and the lines are
well separated.  However, the spectral region around H$\alpha$ required special care because, at the resolution of the G750L grating, \nii\ $\lambda 6548$, H$\alpha$, and \nii\ $\lambda 6584$ are blended and have comparable strengths.  To measure these line fluxes, we fixed the strength of \nii\ $\lambda 6548$ to be one-third that of \nii\ $\lambda 6584$ \citep{Osterbrock2005}, simultaneously fit all three lines using Gaussian profiles of 
width 13~\AA\ (the resolution of the spectrograph in the vicinity of H$\alpha$), and then binned the 
resultant blend to the pixel-scale of the detector.  The result is shown in the inset in Figure~\ref{fig:full_spec}.  Despite being a high-excitation object with bright \oiii\ and undetectable [\ion{O}{2}], the PN has a \nii\ $\lambda 6584$ line that is almost as strong as H$\alpha$.  This
demonstrates that the nebula is very rich in nitrogen,  suggesting that it is a 
Type~I PN \citep{Peimbert1978}; see below for further discussion.

\subsection{Absolute Flux Calibration}
\label{subsec:absolute}

Table~\ref{table:lines} lists the measured fluxes of all detected emission lines relative to H$\beta$, $F/F\rm(H\beta)$, and upper limits for a few undetected lines of interest.\footnote{Table~\ref{table:lines} includes measurements of two weak lines of [\ion{Ne}{3}] and [\ion{S}{3}], detected at low SNR near the ends of the spectra. The resulting abundances will be accordingly uncertain.}  To convert these flux ratios to absolute line fluxes, we normalized to the absolute flux in \oiii\ $\lambda 5007$, which is by far the brightest feature in the spectrum.  We have two independent methods for determining the absolute flux of $\lambda$5007.  The first is through photometry of the PN's \oiii\ emission as recorded in the ACS F502N direct image (\S\ref{sec:cluster}).  Correcting to an infinite aperture, we find that the PN's flux through this filter corresponds to an AB magnitude of $m_{\rm AB} = 19.86$.  If we assume that the PN's continuum emission in the optical is negligible and use the procedures for emission-line photometry outlined by \citet{Jacoby1987}, then this value, combined with knowledge of the F502N filter's transmission curve,\footnote{\url{http://www.stsci.edu/hst/acs/analysis/throughputs}} translates into a monochromatic $\lambda 5007$ flux of $F_{5007} = 2.8 \times 10^{-15}$~erg~cm$^{-2}$~s$^{-1}$, equivalent to an \oiii\ magnitude of $m_{5007} = 22.64$.  

\begin{deluxetable*}{llccc}
\tablewidth{0pt}
\tablecaption{Measured, Dereddened, and Predicted Emission-Line Fluxes}
\tablehead{
\colhead{Wavelength [\AA]}
& \colhead{Ion}
& \colhead{$F/F(\rm H\beta)$\tablenotemark{a}}
& \colhead{$I/I(\rm H\beta)$\tablenotemark{b}}
& \colhead{Model\tablenotemark{c}} }
\startdata
1548+1550 & \ion{C}{4}  &   $<\!0.5$     & $<\!9.1$    &  0.16, 1.89 \\
1906+1909 & \ion{C}{3}] &   $<\!0.5$     & $<\!9.5$    &  2.85, 3.61 \\
3727 & [\ion{O}{2}]   & $<\! 2.7$     & $<\! 5.6$    &  1.79, 2.13 \\ 
3869 & \neiii      & $0.92\pm0.24$  & $1.79\pm0.47$  &  1.79 \\
4363 &\oiii        & $<\! 0.25$     & $<\! 0.35$     &  0.19, 0.25 \\
4686 & \ion{He}{2} & $<\!0.30$       & $<\!0.34$       &  0.00, 0.31 \\
4861 & H$\beta$    & $1.00\pm0.05$  & $1.00\pm0.05$  &  1.00 \\
4959 & \oiii       & $4.82\pm0.40$  & $4.54\pm0.36$  &  4.62 \\
5007\tablenotemark{d} & \oiii       & $15.02\pm0.64$ & $13.78\pm0.60$ & 13.78 \\
5876 & \ion{He}{1} & $<\!0.3$       & $<\!0.2$       &  0.22, 0.18 \\
6548 & \nii        & $2.05\pm0.11$  & $1.04\pm0.07$  &  0.94 \\
6563 & H$\alpha$   & $5.61\pm0.39$  & $2.85\pm0.20$  &  2.80 \\
6583 & \nii        & $5.46\pm0.44$  & $2.75\pm0.22$  &  2.75 \\
6717+6731 & \sii     & $<\!0.3$       & $<\!0.15$      &  0.12, 0.19 \\
9531 & \siii       & $1.63\pm0.81$  & $0.38\pm0.19$  &  0.38 \\
\enddata
\label{table:lines}
\tablenotetext{a}{Observed flux ratios (or upper limits) relative to H$\beta$; uncertainties include the error in H$\beta$ as determined by measurement differences between two of the authors (BDD and GHJ).}
\tablenotetext{b}{Flux ratios (or upper limits) corrected for extinction using a logarithmic extinction at H$\beta$ of $c=0.99 \pm 0.06$; the uncertainties include those of the individual line measurements (including H$\beta$) and the error in the extinction correction.}
\tablenotetext{c}{Predicted flux ratios from the two models described in \S\ref{sec:cloudy} for the detected lines; for the undetected lines, the predictions of both Model 1 and Model 2 are given.}
\tablenotetext{d}{After correction for circumnebular extinction of $c = 0.99$, the measured absolute line flux of \oiii\  $\lambda 5007$ is $1.70 \times 10^{36}$ erg~s$^{-1}$.}
\end{deluxetable*}

Alternatively, we can simply adopt the absolute calibration of the STIS spectrograph, which gives $m_{5007}= 22.91$ ($F_{5007} = 2.18 \times 10^{-15}$~erg~cm$^{-2}$~s$^{-1}$) as the flux of the \oiii\ $\lambda 5007$ line.  Due to the possibility of slit losses, we regard this as a robust lower limit to the line flux.

Both of these estimates are fainter than the value of 
$m_{5007} = 22.30$ derived from the PN.S counter-dispersed imaging reported by M06. 
However, such an offset is consistent with the spectroscopy of C09, which shows that the \oiii\ $\lambda 5007$ equivalent width in the integrated spectrum of the star cluster plus PN is ${\gtrsim}19$~\AA\null.  If the PN.S reductions did not account for the superposed light of the cluster, then, at the nominal
resolution of the instrument \citep{Douglas2002}, flux from the underlying cluster would have boosted the PN's magnitude by ${\gtrsim}0.2$~mag.  As the photometric precision of the M06 survey is $\sim\!15\%$, this correction would make their measurement consistent with that from the ACS.

Given the possibility of systematic errors associated with the PN.S and STIS long-slit data, we adopt the ACS value of $2.8 \times 10^{-15}$~erg~cm$^{-2}$~s$^{-1}$ as the flux of the PN's \oiii\ $\lambda 5007$ emission, and scale the other emission lines to this flux. 

At the distance (750~kpc) and foreground Galactic reddening \citep[$A_V = 0.17$;][]{Schlafly2011} of M31, the $m_{5007}$ magnitude places the PN about 2.5~mag down the \oiii\ luminosity function, as anticipated in \S4.2. The PN emits roughly $60 \, L_{\odot}$ in the observed \oiii\ $\lambda 5007$ emission line. When corrected for circumnebular extinction via a \citet{Cardelli1989} reddening law and the observed H$\alpha$/H$\beta$ ratio (which implies $A_{5007} = 2.38$; see \S\ref{subsec:cloudyconstraints}), the true \oiii\ $\lambda 5007$ luminosity becomes $\sim\!440 \, L_{\odot}$. The remaining columns in Table~\ref{table:lines} will be discussed below.


\section{Fitting the Cluster CMD}


To determine the age of the M31 cluster B477-D075, and thus the progenitor mass of the PN's central star, we used photometry of individual stars in the ACS F814W and F475W filters, as measured and tabulated\footnote{\url{https://archive.stsci.edu/prepds/phat/}} by the PHAT survey team \citep{Dalcanton2012}. We defined our cluster region to be an annulus centered on the cluster, with radii $\phi$ in the range $0\farcs 6 < \phi < 1\farcs 9$ ($2 \lesssim r \lesssim 7$~pc). We avoided the cluster's central core ($\phi < 0 \farcs 6$) due to excessive crowding, and truncated the analysis at $1\farcs9$, beyond which there are few cluster members.

\subsection{Field-Star Removal}

\begin{figure}
\centering
\includegraphics[width=0.473\textwidth]{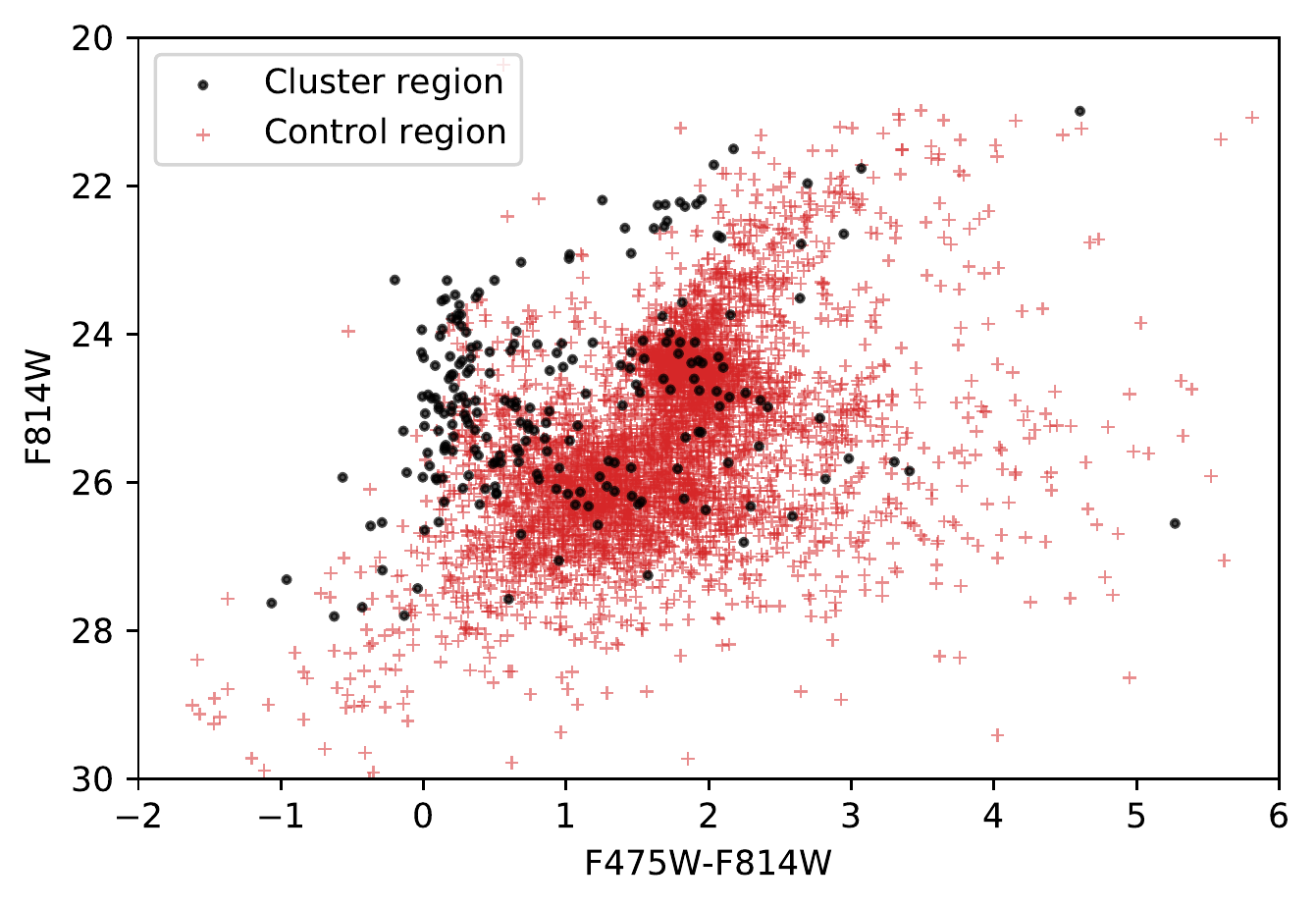}
\caption{Color-magnitude diagrams for the region enclosing M31 cluster B477-D075 (black points) and the control field used for statistical background subtraction (red points). Even without field subtraction, the upper main sequence and turnoff of the cluster are clearly defined.}
\label{fig:fields}
\end{figure}

In order to age-date the cluster, we first statistically removed field stars superposed on the cluster. We defined a control field using an annulus with inner and outer radii of $\phi=4$\arcsec\ and 10\arcsec\ from the cluster center.  (By keeping the control region this close to the cluster, we minimized the effects of differential reddening across the field.)  We then statistically removed stars from the cluster region that lie close to objects in the control field in color-magnitude space, using the following process:

\begin{enumerate}
    \item Calculate the ratio of cluster area to control area, $R = A_{\rm{cluster}}/A_{\rm{control}}$.
    \item For each field star in the control region, generate a uniformly distributed random number, $N$, on the interval [0,1].
    \item If $N<R$, calculate the separation in color-magnitude space, $s$, between the control star and each cluster star using the metric 
    $$s = \sqrt{\left[\Delta (m_{475} - m_{814})\right]^2 + \left[\Delta m_{814}/5\right]^2}$$
    \item If the cluster star with the smallest separation has $s < 0.55$~mag, remove it from the cluster sample.
\end{enumerate}
The limit of $s = 0.55$~mag protects against spurious subtractions due to outliers, while the factor of five in the metric calculation is the same as that used by \citet{Sanner1999, Sanner2000, Sanner2001} for the analysis of open clusters in the Milky Way.  (As in the Sanner et al.\ studies, we find our results are insensitive to the precise value of this denominator.)  Figure~\ref{fig:fields} shows the CMDs for the cluster region and the control field. 

\subsection{Age, Metallicity, and PN Progenitor Mass}
\label{subsec:age-z-mass}

The next step is to fit the statistically cleaned cluster CMD with model stellar isochrones. We used a two-parameter (age and metallicity) grid of isochrones provided in the ACS filters by the MESA Isochrones and Stellar Tracks \citep[MIST, version 1.2;][]{Dotter2016,Choi2016} website\footnote{\url{http://waps.cfa.harvard.edu/MIST/}} and its web interpolator. We adopted MIST's default stellar rotation of $v/v_{\rm crit} = 0.4$, which is  appropriate for stars in the mass range expected for B477-D075's turnoff. (For more on the effects of rotation, see \citealt{Cummings2018}.) Again assuming an M31 distance of 750~kpc, we iterated as follows:

\begin{enumerate}
    \item To estimate the amount of extinction, foreground Galactic plus internal to M31, we adopted a \citet{Cardelli1989} reddening law with $R_V = 3.1$, and chose an initial value for $A_V$ that roughly fit the main sequence of the CMD.
    \item With the value of $A_V$ held constant, we performed a $\chi^2$ fit for each isochrone in our grid, to find the metallicity and age that best fit the data.
    \item We then held the best-fit values of age and metallicity constant and repeated the $\chi^2$ analysis, this time varying $A_V$ until its best-fit value was found.
    \item Using the updated value for $A_V$, we repeated steps 2 and 3 until all three parameters converged.
\end{enumerate}

\begin{figure}
\centering
\includegraphics[width=0.473\textwidth]{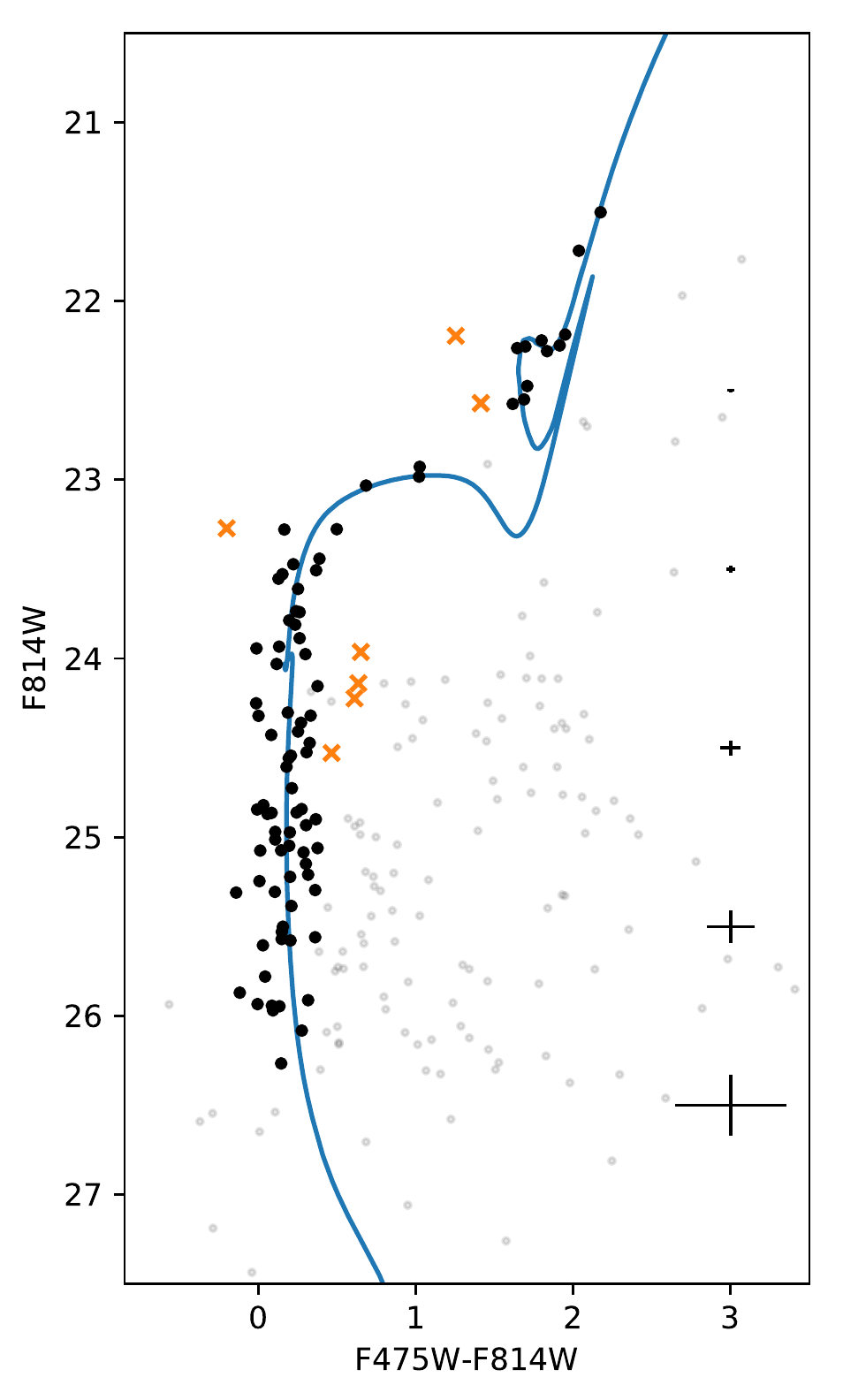}
\caption{Color-magnitude diagram for the M31 cluster B477-D075 in the ACS F475W and F814W filters. Light gray points represent statistically removed field stars, black points show cluster stars that remained after statistical subtraction of the field, and orange crosses show the outliers that were removed manually.  The mean photometric uncertainties quoted by the PHAT team are shown for each magnitude bin in the F814W filter. At an M31 distance of $750\ \rm{kpc}$, the CMD is best fit with a MIST isochrone (blue line) of age $290 \, \rm{Myr}$, with a visual extinction of $A_V = 0.53$~mag, and a metallicity of $Z = 0.0071$.}
\label{fig:cmd}
\end{figure}

During this procedure, we noted that a few of the stars that survived the statistical field-subtraction process had CMD positions inconsistent with the bulk of cluster members.  To deal with these outliers, we examined the CMD during the iteration process and manually removed those stars whose individual $\chi^2$ values made cluster membership extremely unlikely. Figure~\ref{fig:cmd} displays these stars, along with the field-subtracted CMD, and our final best-fitting isochrone. 

Based on this analysis, we find values for the extinction, metallicity, and age of the cluster of $A_V = 0.53$~mag [corresponding to $E(B-V)=0.17$], $Z = 0.0071$, and $\tau = 290 \, \rm{Myr}$. From the resultant $\chi^2$ values, 68\% of the likelihood ($1 \, \sigma)$ lies between $0.52$ and $0.54$~mag in extinction, $0.0068$ and $0.0073$ in metallicity, and 285 and 294~Myr in age. Since the MIST isochrones adopt a proto-solar metallicity of $Z=0.0142$ \citep[from][]{Asplund2009}, our result implies that the cluster is slightly metal-poor, with $\rm[Z/H]=-0.30$. For our adopted parameters, the MIST isochrones indicate the initial mass for post-AGB stars in the cluster to be $3.38^{+0.03}_{-0.02} \, M_{\odot}$.  

As a check on the sensitivity to our choice of theoretical models, we repeated our analysis using the PARSEC isochrones \citep{Bressan2012}, obtained from the CMD 3.2 web  interface.\footnote{\url{http://stev.oapd.inaf.it/cmd}} The $\chi^2$ statistics for fits using these models (which assume no rotation) proved to be significantly larger than those found using MIST tracks, but the overall results were similar.  These fits gave the exact same extinction of $A_V = 0.53 \pm 0.01$, a slightly older cluster age of $327^{+4}_{-6} \, \rm{Myr}$, and a slightly lower cluster metallicity of $Z = 0.0069^{+0.0002}_{-0.0003}$.  With these models, the inferred post-AGB progenitor mass is $3.19^{+0.03}_{-0.02} \, M_{\odot}$.

For both sets of theoretical isochrones, our derived cluster age is in general agreement with the ages determined by \citet{Kang2012} and C09, and only slightly older than those derived by \citet{Johnson2016} and \citet{deMeulenaer2017}. Additionally, both isochrone models yield a foreground plus local M31 extinction of $A_V = 0.53 \pm 0.01$, which lies near the middle of the range of values found by other authors. Finally, both sets of isochrones give PN progenitor masses in the range of 3.2--$3.4 \, M_{\odot}$. For the rest of our discussion, we will adopt the results from the MIST isochrones.

\section{Photoionization Modeling of the PN}
\label{sec:cloudy}

\subsection{Rationale} 
\label{subsec:cloudyrationale}

We now turn to constructing a photoionization model for the PN\null. Our primary goal is to estimate the chemical abundances in the nebula, particularly that of nitrogen. This will allow a comparison with  predictions of AGB evolutionary codes \citep[e.g.,][and MIST]{Cristallo2015, Karakas2016, Marigo2017}.  Photoionization modeling provides a way to derive or constrain the properties of ionized nebulae when the observational information is insufficient for a direct measurement. For the PN M31~B477-1, there are insufficient constraints to derive any abundances directly, for example via ionization correction factors \citep[e.g.,][]{Kingsburgh1994, DelgadoInglada2019}.  In this situation, a modeling program like \cloudy\  \citep{Ferland2013} provides an effective alternative.

For bright, well-observed PNe, a \cloudy\ model can be constrained by a large number of observables, such as (position-dependent) emission-line strengths, the nebula's absolute line fluxes and angular diameter, and the central star's luminosity.  One adjusts a variety of input parameters, such as the abundances of key elements (helium, carbon, nitrogen, oxygen, neon, sulfur, and argon), the nebula's physical radius and electron density, and the central star's temperature and luminosity, to create a model whose observables closely match the constraints.  Central-star mass is then derived from the exciting star's luminosity and temperature via comparisons with models of post-AGB stellar evolution.  Finally, the mass of the progenitor star is estimated using an initial-mass\slash final-mass relation (IFMR), which comes either from models \citep[e.g.,][hereafter B16]{Bertolami2016} or empirical data \citep[e.g.,][]{Cummings2018, El-Badry2018}.

For faint extragalactic PNe that are challenging to observe, the nebula is unresolved, the central star is invisible, and there are relatively few detectable emission lines.  Consequently, photoionization models will not provide unique solutions.  Nevertheless, the parameter space can still be sufficiently delimited to provide interesting insights into the object's chemical composition, luminosity, and evolutionary status.

\subsection{Input Constraints to \cloudy\ } \label{subsec:cloudyconstraints}

The measured line fluxes for the PN M31~B477-1 are listed in column~3 of Table~\ref{table:lines}. These must first be corrected for extinction. We estimated this from the Balmer decrement, $F({\rm H\alpha)}/F({\rm H\beta})$, by assuming a standard (Case~B) unreddened value of 2.85 \citep{Brocklehurst1971}, and using the \citet{Cardelli1989} extinction law with $R_V = 3.1$. This calculation yields a logarithmic H$\beta$ extinction of $c = 0.99 \pm 0.06$ [corresponding to $E(B-V)=0.68$ or $A_V=2.13$]. Based on this value, we give the dereddened line ratios (or upper limits) in column~4 of Table~\ref{table:lines}, denoted $I/I({\rm H\beta})$.  Note that the quoted errors for these ratios are driven primarily by the uncertainty in the continuum level at H$\beta$, as the H$\beta$ flux directly affects all the observed line ratios (which are relative to H$\beta$), and especially the reddening-corrected line ratios, which are all dependent on the adopted value of~$c$.

As expected, the circumnebular extinction derived from the Balmer decrement is substantially larger than that inferred for the star cluster as a whole. Interestingly, the amount of excess reddening, $\Delta E(B-V) \simeq 0.5$, is near the high end of the range of values found for samples of PNe in M31's bulge \citep{Davis2018} and NGC~5128's envelope \citep{Walsh2012}.  Since the progenitor mass for the B477-D075 PN was likely higher than for the PNe in either of these two older stellar populations, and since the B477-D075 PN is still quite young (based on the density and radius), a larger amount of circumnebular material, and thus a high value of $c$, is not surprising.

\subsection{\cloudy\ Modeling Assumptions} \label{subsec:cloudyassumptions}

We considered two different approaches in our modeling philosophy. For our first analysis, we attempted to derive the properties of the PN independently of any external information, i.e., with no knowledge of the cluster's age and metallicity.  Not surprisingly, the solution space for this approach is very broad due to degeneracies between several of the parameters. Two parameter pairs were particularly troublesome: stellar temperature vs.~oxygen abundance, and stellar luminosity vs.~dust-to-gas ratio. For example, by varying the stellar temperature from 60,000\,K to 130,000\,K, we could build models that span a very large range in the nebular oxygen abundance and central-star luminosity and still satisfy the constraints imposed by the STIS spectroscopy.  One constant, however, is the result for nitrogen: all of our models yielded values for the nitrogen abundance that are significantly greater than the cluster's initial value.  This result is not surprising: the strength of \nii\ $\lambda\lambda 6548,6583$, combined with the strong \oiii\ $\lambda 5007$ and the weak lines of the low-ionization species other than nitrogen (e.g., O$^{+}$, S$^{+}$), are highly indicative of nitrogen enhancement.

After performing this initial survey of parameter space, we then took a second approach to modeling and used the fact that the PN is very probably a member of the open cluster B477-D075 (see \S\ref{sec:isitamember}).  This allowed us to constrain two key parameters:


1)~Metallicity: we fixed the abundances for carbon and oxygen using the central star's post-AGB surface composition, as derived from the predictions of a MIST evolutionary track for a $3.38\,M_\odot$ star with $Z = 0.0071$.
Nitrogen, neon, and sulfur abundances were then adjusted to allow for the best match to the observational data for those elements.  We note that nitrogen, in particular, is known to be enhanced during the stellar evolution of intermediate-mass stars \citep[e.g.,][hereafter H18, and references therein]{Henry2018}.  Carbon can also be modified in the progenitor, but we have no information on that element, other than our upper limits on \ion{C}{4} and \ion{C}{3}], which were never violated. We return to these assumptions in \S\ref{subsec:cloudycautions}.

2)~Central-star mass and luminosity: As discussed in \S5.2, the MIST isochrone for our estimates of the cluster's age and metallicity implies an initial mass of $\sim\!3.4\, M_\odot$ for the progenitor star. Based on the IFMR and post-AGB evolutionary tracks of B16 (or the predictions from MIST), the mass of the central star should be $0.78\,M_\odot$, and the central star luminosity should range from $\sim\!14{,}500$ $L_{\odot}$ for low-temperature central stars to $\sim\!13{,}000$ $L_{\odot}$ for high-temperature stars.


With these additional constraints, the \cloudy\ models became highly constrained, principally by the absolute luminosity of \oiii\ $\lambda 5007$ (see \S\ref{subsec:absolute}) and the flux ratio of $I(5007)/I(\rm H\beta$). 

\subsection{Deriving a \cloudy\  Model} \label{subsec:cloudysteps}

For typical PN central-star temperatures (e.g., between 65,000\,K and 120,000\,K), the high central-star luminosity (\S\ref{subsec:cloudyassumptions}) causes the PN's \oiii\  emission-line luminosity and $I(5007)/I({\rm H}\beta$) ratio to exceed the observed values. By either lowering or raising the star's temperature beyond this range, we can manipulate the ionization balance so as to reduce the fraction of O$^{++}$ ions, and thereby reduce the strength of $\lambda 5007$.  We can also reduce the PN's line-luminosity by introducing dust grains into the nebula, as this would shift more of the central star's energy into the mid- and far-IR\null.

On this basis, we created two models for the PN, Models 1 and 2, which represent the solutions for low and high central-star temperatures, respectively.  To approximate the radiation field of the central star, 
we used the non-LTE model stellar atmospheres of \citet{Rauch2003} for high-gravity stars. Specifically, given the expected high mass for our central star, we adopted a surface gravity of $\log g = 6.5$ \citep[e.g.,][]{Balick2013}.  However, the exact value of this parameter has little influence on the \cloudy\  models.

The steps involved in developing and finalizing the models are summarized as follows:

1) We adopt initial values for the chemical abundances, focusing on the primary coolants: He, C, N, O, Ne, S, and Ar. As described above, we assume nebular abundances derived from the MIST predictions for the surface composition of a post-AGB central star with the metallicity and age of the cluster. However, we adopt an enhanced helium abundance by number of $\rm He/H=0.14$, which is typical for a Type~I PN \citep{Peimbert1983}, under the assumption (based on the strength of [\ion{N}{2}]) that our PN falls into that class.

2) We assign a nebular electron density of 1800\,cm$^{-3}$. This value is roughly consistent with the density derived from the [\ion{S}{2}] $\lambda6717/\lambda6731$ ratio seen in the higher-resolution cluster spectrum of C09. (In their data, $\lambda 6717$ and $\lambda 6731$ are roughly equal, implying $n_e \simeq 900$\,cm$^{-3}$; \citealp{Osterbrock2005}.)  We note that because the C09 spectrum comes from a ground-based fiber-fed instrument with no local sky subtraction, the line ratios of low-ionization features, such as those of [\ion{S}{2}], may be contaminated by emission from M31's low-density diffuse interstellar gas \citep[e.g.,][]{Greenawalt1997, Galarza1999}. (The [\ion{S}{2}] lines in our STIS spectrum are, at best, marginally detected.) We therefore increased the adopted density above the value indicated by C09's [\ion{S}{2}] ratio.  This is justified by our expectation that a high-mass central star will have a very rapid evolutionary timescale, leaving little time for the ionized ejecta to expand into a low-density nebula.  In any case, the impact of a small error in density is minor, as this parameter primarily affects the derived nebular radius and ionized mass, rather than the elemental abundances. See \S\ref{subsec:cloudycautions} for further discussion.

3) We then vary the central-star temperature to achieve a match for the dereddened line ratio of $I(5007)/I(\rm H\beta)$. We find that there are two central-star temperature regimes where we can accomplish this:  low-temperature (Model~1), and  high-temperature (Model~2). Both yield a nebular ionization balance that produces the observed line ratios; intermediate-temperature models yield $I(5007)/I({\rm H}\beta$) ratios that exceed this constraint.  If we had not pre-defined the oxygen abundance, then the derived stellar temperature could be anywhere between $\sim$55,000\,K to $\sim$125,000\,K, with the low end of the range defined by the upper limit on \ion{He}{1}, and the high end delimited by the limit on \ion{He}{2}.
    
4) We tune the dust-to-gas ratio to match the luminosity of \oiii\ $\lambda 5007$ at the stellar temperature found in step 3.  This tuning is done iteratively with the previous step until both constraints are satisfied.

5) We adjust the remaining chemical abundances to match the line ratios in Table~\ref{table:lines}.  The most important element in the table is nitrogen; constraints on the other emission lines are relatively weak, except perhaps for Ne/H and S/H\null.  In principle, we could also have adjusted the carbon abundance to help moderate the PN electron temperature and reduce the pressure on the \oiii\  luminosity and line ratio.  However, in a Type~I PN, carbon is more likely to be depleted than enhanced \citep{Kaler1990, Karakas2014}. Unfortunately, our limits on \ion{C}{4} $\lambda 1550$ and \ion{C}{3}] $\lambda 1909$ from the STIS spectra are too weak to effectively constrain the carbon abundance.

6) We iterate Steps 3, 4, and 5 until we reach an acceptable match to all the observed line fluxes.  

Table~\ref{table:PNproperties} summarizes the properties of our two final \cloudy\ models. For both models, the table gives the adopted parameters of the PN, and the chemical abundances of six species (on the usual scale of 12 plus the logarithm of the abundance by number relative to hydrogen). As noted in the discussion above, the abundances of He, C, and O were adopted in advance; only N, Ne, and S were derived from our observations. Based on the nebular parameters and chemical abundances, we predict the emission-line fluxes for the two models, and give them in the final column of Table~\ref{table:lines}. For the detected lines, both models match the observations quite precisely. For the undetected lines, the final column gives the predicted fluxes from the two models. Note that both models marginally satisfy the upper limits on the fluxes of \ion{He}{1} $\lambda 5876$ (Model 1) and \ion{He}{2} $\lambda 4686$ (Model 2), at the temperatures of the exciting star. We discuss this concern further in \S\ref{subsec:cloudycautions}.
    
\subsection{Results from the \cloudy\  Models} \label{subsec:cloudyresults}

Table~\ref{table:PNproperties} gives nitrogen-to-hydrogen ratios of $12+\log({\rm N/H})=8.36$ and 8.25, for the cool and hot central-star temperatures, respectively. Both values exceed the solar nitrogen logarithmic abundance of 7.83 (by 0.53 and 0.42~dex), and are even more elevated when compared to the cluster's primordial abundance of $\rm[Z/H] = -0.3$ (or $12+\log({\rm N/H})=7.53$) derived in \S\ref{subsec:age-z-mass}. Relative to this initial value, the N/H enhancement factors for our two models are 6.2 and 4.8.  This result is the primary outcome of our photoionization modeling, and, as described in \S\ref{subsec:cloudyassumptions}, it is highly robust to variations across a wide range of input assumptions.

\begin{deluxetable*}{lccl}
\tablewidth{0pt}
\tablecaption{Model PN Parameters and Element Abundances from \cloudy\ v17.01 Analysis\label{table:PNproperties}}
\tablehead{
\colhead{Parameter}
& \colhead{Model 1\tablenotemark{a}}
& \colhead{Model 2\tablenotemark{b}}
& \colhead{Remarks}}
\startdata
Central-star mass [$M_{\odot}$]     & $0.78\pm0.03$   & $0.78\pm0.03$   & From cluster age, $Z$, and MIST prediction \\
Stellar luminosity [$L_{\odot}$]    & 14,500          & 13,000          & From central-star mass and MIST evolutionary track \\
Stellar temperature, $T_\bigstar$ [K]      & $59{,}800$      & $126{,}300$     & Adjusted to match $I(5007)/I({\rm H}\beta$) \\
$12+\log($He/H)                     & 11.15 (10.98)   & 11.15           & Set to typical Type I PNe composition\tablenotemark{c} \\
$12+\log($C/H)                      & 8.32 (8.17)     & 8.32            & Set to post-AGB value from MIST evolutionary track \\
$12+\log($N/H)                      & 8.36 (7.57)     & 8.25            & Derived from $I(6583)/I({\rm H}\beta$) \\
$12+\log($O/H)                      & 8.42 (8.43)     & 8.42            & Set to post-AGB value from MIST evolutionary track \\
$12+\log($Ne/H)                     & 8.01 (7.67)     & 7.97            & Derived from $I(3869)/I({\rm H}\beta$) \\
$12+\log($S/H)                      & 6.39 (6.86)     & 6.46            & Derived from $I(9531)/I({\rm H}\beta$) \\
Electron density [cm$^{-3}$]        & 1,800           & 1,800           & Fixed; see text \S6.4 \\
Electron temperature [K]            & 12,400          & 14,200          & Depends on $T_\bigstar$ and abundances \\
Nebular radius [pc]                         & 0.153           & 0.156           & Radius of ionized nebula, depends on density \\
Nebular ionized mass [$M_{\odot}$]   & 0.40           & 0.42        & Inversely proportional to density \\
$c$ (log extinction at H$\beta$)    & $0.99\pm0.06$   & $0.99\pm0.06$   & From $F({\rm H}\alpha)/F({\rm H}\beta$) and Case~B assumption \\
Dust-to-gas mass ratio\tablenotemark{d}  & 0.013           & 0.015           & Adjusted to match absolute $\lambda$5007 luminosity \\
\enddata
\tablenotetext{a}{Abundance values are by number; values in parentheses are the original composition of the PN progenitor, derived by adopting the protosolar values from \citet{Asplund2009} and reducing them by 0.3~dex (except for He) to match the subsolar metallicity of the host cluster. Both models
adopt C and O abundances for the PN from the post-AGB MIST tracks, and the He abundance for a typical Type~I PN\null. Model~1 represents a PN whose central star has a \citet{Rauch2003} model atmosphere at the stellar temperature given and $\log g = 6.5$. } 
\tablenotetext{b}{For Model 2, the stellar temperature was adjusted beyond the peak of the $I(5007)/I({\rm H}\beta)$  ratio to again match
the observed data. \citet{Rauch2003} atmosphere models were again adopted.}
\tablenotetext{c}{From \citet{Peimbert1983}}
\tablenotetext{d}{Using the \cloudy\ default grain composition}
\end{deluxetable*}

Our results also strongly suggest that the B477-D075 PN is optically thick, with significant amounts of neutral material surrounding the ionized nebula.  If this were not the case, the outer zone of the planetary, where most of the \nii\ emission is produced, would be truncated, and the strong \nii\ lines seen in Figure~\ref{fig:full_spec} would have to come from the nebula's high-excitation inner regions.  That would require a {\it much\/} greater nitrogen abundance than the already high values quoted in Table~\ref{table:PNproperties}. 

An additional argument for the existence of circumnebular neutral material comes from the mass budget.  As first shown by \citet{Stromgren1939}, the total ionized mass of an optically thick \ion{H}{2} region (or PN) is inversely proportional to its density. For the case of M31~B477-1, the ionized mass is about $0.4 \, (1800 \, {\rm cm}^{-3}/ n_e) \, M_{\odot}$ (see Table~\ref{table:PNproperties}).   Since the initial mass of the PN's progenitor was $3.38 \, M_{\odot}$ and the central star's final mass is $\sim\! 0.78 \, M_{\odot}$, more than $2\,M_\odot$ of material is unaccounted for. There must therefore be a large amount of dust and neutral gas surrounding the ionized central zone.  The fact that the PN is affected by $A_V \simeq 1.6$~mag more extinction than its fellow cluster stars supports this conclusion.  

Our discussion has wider applicability. The IFMR suggests that stars with initial masses of $\gtrsim$$3 \, M_{\odot}$ lose 
$\gtrsim$$2 \, M_{\odot}$ of material during the AGB phase, and their luminous post-AGB cores have evolutionary timescales that do not allow much time for this matter to disperse.  As a result, the ionized zones of bright PNe should be small, dense, and surrounded by large amounts of dust and neutral gas. This picture is supported observationally by \citet{Davis2018}, who concluded that substantial amounts of circumnebular extinction are common in \oiii-luminous PNe in the bulge of M31, the LMC, and giant elliptical galaxies.

\subsection{Cautionary Notes} \label{subsec:cloudycautions}

We have already noted that our photoionization models are not unique, and are somewhat dependent on several assumptions. In particular, the oxygen abundance used for our modeling is based on the assumption that the PN is a member of the M31 open cluster B477-D075.  Our determination that there is a nitrogen enhancement, however, is robust: even without the cluster prior, we could not adjust N/H by more than $\pm 0.15$~dex.  No model resulted in a nitrogen abundance that is close to solar, let alone to the subsolar initial content in the cluster.

Other aspects of the modeling that may result in  minor variations in the results include the following:

1)  Aside from varying nitrogen, we did not explore the impact of changing the abundances of individual elements. For neon and sulfur, we tuned the abundances so that the line ratios were matched. For carbon, we fixed the abundance to the MIST value for post-AGB stars with initial an metallicity matching that of the cluster.  We could make a physical argument for decreasing carbon, as C should be converted to N during CNO burning.  However, because carbon is an important nebula coolant, any reduction in its abundance would need to be offset by other coolants in order to avoid pushing \oiii\ $\lambda 5007$ over the observed limit.  We struggled with high values of \oiii\ throughout, so reducing carbon would further exacerbate the situation; the best way to compensate for carbon reduction is to decrease the central-star temperature in Model 1 or increase it in Model 2 (see next item).

2) The low-temperature model (Model~1) meets all the observational constraints. However, it has a central-star temperature (59,800\,K) that is uncomfortably low for a bright, high-excitation PN\null. This condition becomes exacerbated if the electron density is much higher than the adopted value (see next item).

3) As noted in \S\ref{subsec:cloudysteps}, we do not have a strong constraint on the nebular electron density. To assess the impact that density has on our results, we computed models at half and double the adopted value of $1800\,\rm cm^{-3}$. At higher densities, the low-temperature models are driven to even lower temperatures (by $\sim$1000\,K for $3600\,\rm cm^{-3}$); the opposite is true for the high-temperature models.  The effect for nitrogen is to (slightly) increase as electron density increases and vice versa. For example, a low-temperature model with $n_e = 3600\,\rm cm^{-3}$ has a nitrogen abundance that is 0.09~dex larger than the 12 + $\log({\rm N/H}) = 8.36$ value found at $1800\,\rm cm^{-3}$, and 0.08 dex lower when $n_e = 900\,\rm cm^{-3}$.

4) We did not consider changing the grain composition or size distribution from the defaults provided by \cloudy\null. Doing so represents an exercise beyond the scope of this paper, but the result would likely be a lower \oiii\ flux and models that are more tolerant of the observed PN line ratios.  A larger grain size, in particular, would lead to less collisional heating via the photoelectric effect and allow our Model~1 to converge at a higher central-star temperature.  For the PN in B477-D075, a variation in the grain properties is certainly plausible:  the dust-to-gas ratio in Galactic PNe is known to vary by more than a factor of 10, with one viable explanation being a correlation between the fraction of dust within a nebula and central-star mass \citep{Stasinska1999, Ciardullo1999}.  Since the IFMR suggests that M31~B477-1 has a high-mass core, a larger dust content within the nebula (and a large amount of circumnebular extinction) is reasonable.

\section{Discussion} \label{sec:discussion}

\subsection{M31 B477-1 as a Type I PN}
\label{subsec:TypeI-PN}
Type~I PNe were defined by \citet{Peimbert1978} as a relatively small subset of
planetaries that have a high abundances of helium and nitrogen. Based on their
kinematics and Galactic distribution, it is likely that Type~I PNe arise from 
more-massive progenitors than the general PN population \citep{Peimbert1983}.


Our abundance analysis of M31~B477-1 shows that it has at least some of the
characteristics of Type~I PNe: in particular, it is  overabundant in nitrogen.  This robust result can be seen immediately from our spectra
(Figure~\ref{fig:full_spec}), in which the low-ionization lines of 
[\ion{O}{2}] $\lambda 3727$ and \sii\ $\lambda\lambda 6717,6731$ are at best, barely visible, but \nii\
$\lambda 6583$, which has a similar ionization potential, is as bright as H$\alpha$.  Our models give a nitrogen abundance that is ${\sim}6$ times that expected from the modestly sub-solar metallicity of the host cluster. 

In our modeling, we assumed that M31~B477-1 is a Type~I PN with an enhanced abundance of helium as well as nitrogen.  While our STIS spectra can only place upper limits on the key helium lines of \ion{He}{1} $\lambda 5876$ and \ion{He}{2} $\lambda 4686$, a careful examination of C09's Hectospec integrated spectrum appears to show weak \ion{He}{1} $\lambda 5876$ abutting the stellar absorption lines from the sodium doublet. Taken at face value, the strength of this \ion{He}{1} detection is consistent with a significant helium enhancement.  However, given the uncertain flux calibration of the ground-based spectrum and low signal-to-noise of the detection, we chose not to use this line as a constraint in our modeling. Nevertheless, it does suggest a helium abundance consistent with that of a Type~I PN and the predictions of either Model~1 or Model~2 from \cloudy.

From our isochrone analysis of the host cluster, the PN progenitor star must have been relatively massive, $\sim$$3.4 \, M_\odot$, and slightly metal-poor, with $Z \simeq 0.007$.  Both values are consistent with previous literature discussions of the cluster (see Table~\ref{table:clusterdata}): age estimates for B477-D075 from SED and CMD analyses generally yield turnoff masses between $\sim$3.22 and $\sim$$3.88 \, M_{\odot}$, while integrated-light spectroscopy implies $Z = 0.0067$ \citep{Chen2016}.  
Thus, we find reasonably secure values for the mass and metallicity of the PN progenitor.

\subsection{The Minimum Mass for Hot-Bottom Burning}

In a recent review of the observed abundances of He, C, and N in PNe, H18 showed that an overabundance of N
is common among PNe. However, the N/O ratio does increase dramatically in PNe whose progenitors are above a certain mass; based on PN statistical distances and comparisons to post-AGB evolutionary tracks, H18 estimated this transition mass to be $\sim\! 2.5 \, M_{\odot}$.
A similar result was found by \citet{Fang2018} for a sample of PNe in M31.

From a theoretical standpoint, it is expected that dredge-up will cause the ejecta of AGB stars to be  nitrogen rich, with N/H ratios that are $\sim$0.2~dex above solar \citep[e.g.,][]{Ventura2015, Karakas2016}.  However, for stars above a certain mass,  the energy generated during the thermally pulsing AGB phase becomes so large that the convective envelope extends all the way down to the hydrogen-burning shell.  As a result of this ``hot-bottom burning" (HBB),
products of the CNO cycle are immediately convected to the surface, where they are subsequently lost into space via a superwind.  The result is a Type~I PN, in which
the nebular abundances of He and N are significantly enhanced, with $N({\rm He})/N({\rm H}) \gtrsim 0.125$ and 
$12 + \log{\rm (N/H)} \gtrsim 8.5$
\citep[e.g.,][]{Peimbert1983}.  

While the existence of a lower-mass limit at which HBB occurs seems clear, the precise mass of its onset remains uncertain. The MONASH \citep{Karakas2014, Karakas2016}, FRUITY \citep{Cristallo2011, Cristallo2015}, and COLIBRI \citep{Marigo2013, Marigo2017} stellar-evolution codes all predict that HBB only occurs in stars with progenitor masses greater than $\sim\!4 \, M_{\odot}$; none come close to creating a $12 + \log({\rm N/H}) \simeq 8.5$ nitrogen abundance in a PN with a $\sim\! 3.4 \, M_{\odot}$ progenitor.  By contrast, models based on the \hbox{LPCODE} \citep{Bertolami2016} and ATON \citep{Ventura2005, DiCriscienzo2016} codes, which incorporate a prescription for overshooting at the boundary of convective cores, have HBB occurring in stars with masses as low as $\sim\! 3 \, M_{\odot}$.   These appear to be in better agreement with our observations, and with the review of H18.

Observational studies of PNe in the Milky Way (H18) and M31 \citep{Fang2018} suggest that HBB can occur in stars with masses perhaps as low as $\sim\! 2 \, M_{\odot}$.  However, it is important to note such estimates are indirect: they rely on knowing the distances to the PNe, modeling the PN spectra to derive the central stars' luminosity and temperature, comparing these numbers to
theoretical post-AGB evolutionary tracks to determine the object's core mass,
and finally translating the core mass into a progenitor mass via an adopted
IFMR. 

Our observations of the PN in B477-D075 provide a more direct determination of the progenitor mass for
a nitrogen-enhanced PN\null.  The STIS spectra clearly show that the PN is overabundant in nitrogen, by a factor of $\sim$5--6 relative to the initial composition. Our CMD analysis implies the mass and metallicity of the progenitor star is $M \simeq 3.4 \, M_{\odot}$ and $Z \simeq 0.007$.  This places a hard upper limit on the minimum mass required for the onset of HBB\null. 

\subsection{Follow-up Studies}

We have argued that the composition of the PN in B477-D075 provides 
direct evidence that HBB occurs in stars of masses lower than generally
predicted in many standard models of stellar evolution.  
Several additional studies would help test our conclusion. 

First, if M31 B477-1 is a Type~I PN, it should also be overabundant in helium.  This could be confirmed by measuring the \ion{He}{1} $\lambda 5876$ and \ion{He}{2} $\lambda4686$ emission lines. Our STIS spectra did not have sufficient signal-to-noise to detect these features, and a deeper exposure with \HST/STIS
would probably be prohibitively costly in observing time.  But \ion{He}{1} and \ion{He}{2} detections, or tighter upper limits, can be obtained with ground-based spectroscopy from a large telescope and careful subtraction of the cluster contamination.  (Indeed, the \ion{He}{1} line is probably present in C09's Hectospec spectrum, as discussed in \S\ref{subsec:TypeI-PN}.)  In doing so, we would also be able to distinguish between our two viable models, which have very different central-star temperatures.

In addition, we have assumed throughout this paper that the PN is in fact a physical member of B477-D075.  The data in hand suggest that this is very likely, but the $\sim\! 10$\,km~s$^{-1}$ uncertainty in our velocity measurements leaves room for doubt. Since the expected velocity dispersion of the cluster is $\sim$1~km~s$^{-1}$, a critical test would be to obtain RVs to this precision for both the PN and the cluster.
Again, this could be done with a large ground-based telescope, preferably with an instrument capable of placing both objects  on the slit simultaneously.

Finally, we have a good measurement of the PN's circumnebular extinction. However, the dust-to-gas ratio within the nebula is essentially unconstrained, and this affects the conversion of ionizing flux to nebular emission.  Our \cloudy\ models predict a far-IR 24~$\mu$m nebular
flux of ${\sim}2 \times 10^{-26}$~erg~s$^{-1}$~cm$^{-2}$~Hz$^{-1}$; this is a factor of $\sim$40 below a 3$\sigma$ noise limit of ${\sim}8 \times
10^{-25}$~erg~s$^{-1}$~cm$^{-2}$~Hz$^{-1}$ that we measure from an archival\footnote{\url{https://sha.ipac.caltech.edu/applications/Spitzer/SHA/}} {\it Spitzer}/MIPS 24~$\mu$m scan of the region (GTO-99: PI G.~Rieke).  However, even though the PN is not detected in the MIPS scan, it should be detectable with the Mid-Infrared Instrument on the James Webb Space Telescope. A combination of deeper optical spectroscopy and mid-IR photometry would fix the the nebula's abundance, along with the central star's temperature and luminosity.  This would allow a more precise determination of the PN's global properties and evolutionary state,  and an even tighter constraint on the stellar masses at which HBB occurs.

\medskip

\acknowledgments

Support for program number GO-14794 was provided by NASA
through grants from the Space Telescope Science Institute, which is
operated by the Association of Universities for Research in Astronomy,
Inc., under NASA contract NAS5-26555. We are grateful to Alison Vick and Molly Peeples for \HST\/ scheduling support, and Charles Proffitt for advice on blind offsets. 
We thank Nelson Caldwell for providing us the Hectospec  spectrum, and Karen Kwitter and Richard Henry for useful discussions.
We also thank Gary Ferland for providing valuable guidance in using \cloudy\null.
The Institute for Gravitation and the Cosmos is supported by the Eberly College of
Science and the Office of the Senior Vice President for Research at The
Pennsylvania State University.

%

\vspace{5mm}
\facilities{HST (STIS, ACS, WFC3), MMT}


\software{\cloudy\ \citep{Ferland2013}}, MIST (\url{http://waps.cfa.harvard.edu/MIST/})

\end{document}